\documentclass[preprint,12pt]{elsarticle}

\usepackage{amssymb,amsmath,amsfonts}
\usepackage{bm} 
\usepackage{mathtools} 
\usepackage{commath} 
\usepackage{ mathrsfs }

\usepackage[per-mode = symbol, exponent-product = \cdot, copy-decimal-marker]{siunitx}
\protected\def\SIpi{\mathnormal\pi}
\usepackage{nomencl}
\makenomenclature

\usepackage{float}
\usepackage{morefloats}

\usepackage{dcolumn,booktabs,multirow,threeparttable}
\usepackage{array}

\usepackage{graphicx} 
\graphicspath{{./Figures/}{./MATLAB/Figures/}} 
\usepackage{tikz}
\usepackage{pgfplots}

\usepackage[font=small,labelfont=bf]{caption}
\usepackage{subcaption}
\usepackage{epstopdf}
\usepackage{amssymb}

\usepackage{makecell}

\biboptions{authoryear}

\usetikzlibrary{calc}

\usepackage{setspace}
\usepackage[nomarkers]{endfloat} 
\usepackage{xcolor}
\usepackage{soul}

\journal{Engineering Fracture Mechanics}

\begin{document}

\nomenclature{$\dot{(~~)}$}{Time derivative}
\nomenclature{$\sigma_{ij}$}{Cauchy stress components}
\nomenclature{$\varepsilon_{ij}$}{Total engineering strain components}
\nomenclature{$\varepsilon^p_{ij}$}{Plastic engineering strain components}
\nomenclature{$r$, $\theta$}{Polar coordinates related to crack tip position}
\nomenclature{$K_I$}{Mode I stress intensity factor}
\nomenclature{$K_{II}$}{Mode II stress intensity factor}
\nomenclature{$f_{ij}$}{Dimensionless mode functions}
\nomenclature{$K_0$}{Reference stress intensity factor}
\nomenclature{$E$}{Young's modulus}
\nomenclature{$E_t$}{Tangent modulus}
\nomenclature{$\nu$}{Poisson's ratio}
\nomenclature{$\Gamma_0$}{Fracture energy}
\nomenclature{$R_0$}{Reference plastic zone size}
\nomenclature{$\sigma_y$}{Initial yield stress}

\nomenclature{$\Phi$}{Traction energy potential}
\nomenclature{$\delta_t$}{Tangential crack separation}
\nomenclature{$\delta_n$}{Normal crack separation}
\nomenclature{$\delta_t^c$}{Critical tangential crack separation}
\nomenclature{$\delta_n^c$}{Critical normal crack separation}
\nomenclature{$\lambda$}{Non-dimensional crack separation}
\nomenclature{$\hat{\sigma}$}{Peak traction}
\nomenclature{$T_n$}{Normal crack face traction}
\nomenclature{$T_t$}{Tangential crack face traction}

\nomenclature{$L_{ijkl}$}{Tensor of instantaneous moduli}
\nomenclature{$F$}{von Mises yield surface}
\nomenclature{$s_{ij}$}{Deviatoric stress}

\nomenclature{$\sigma_e$}{von Mises stress}
\nomenclature{$\beta$}{Material state parameter}
\nomenclature{$\tilde{\sigma}_{ij}$}{Local Cauchy stress components}
\nomenclature{$\alpha_{ij}$}{Back stress components}
\nomenclature{$\dot{\mu}$}{Proportionality coefficient}

\nomenclature{$\tilde{s}_{ij}$}{Local deviatoric stress}
\nomenclature{$f$}{Arbitrary field quantity}
\nomenclature{$\dot{a}$}{Crack propagation speed}
\nomenclature{$x_i$}{Cartesian coordinates}
\nomenclature{$x_i^0$}{Upstream coordinate in history independent region}
\nomenclature{$x_i^*$}{Downstream coordinate}

\nomenclature{$\mathscr{L}_{ijkl}$}{Isotropic elastic stiffness tensor}
\nomenclature{$t_i$}{Surface tractions}
\nomenclature{$V$}{Bounding volume}
\nomenclature{$S$}{Bounding surface}
\nomenclature{$S_c$}{Interface surface}

\nomenclature{$\mathscr{M}_{ijkl}$}{Elastic compliance tensor}
\nomenclature{$u_i$}{Displacement field}
\nomenclature{$\delta_{ij}$}{Kronecker delta}
\nomenclature{$[N]$}{Shape function matrix}
\nomenclature{$[B]$}{Strain-displacement matrix for Q8 elements}
\nomenclature{$[B_c]$}{Strain-displacement matrix for cohesive elements}

\nomenclature{$Q_t$}{Additional degree of freedom in control algorithm}
\nomenclature{$Q_n$}{Additional degree of freedom in control algorithm}
\nomenclature{$\Delta_t$}{Enforced tangential crack tip opening}
\nomenclature{$\Delta_n$}{Enforced normal crack tip opening}
\nomenclature{$N_1$}{Upper crack tip node}
\nomenclature{$N_2$}{Lower crack tip node}
\nomenclature{$N_{BC}$}{Random boundary node}
\nomenclature{$C$}{Far-field scaling factor}
\nomenclature{$K_{ss}$}{Far-field stress intensity factor for steady-state crack growth}
\nomenclature{$L_{e,min}$}{Minimum element length}
\nomenclature{$w$}{Energy dissipation density}
\nomenclature{$\alpha_e$}{Effective back stress}
\nomenclature{$\alpha_{ij}'$}{Deviatoric back stress}

\begin{frontmatter}




\title{Steady-state fracture toughness of elastic-plastic solids: Isotropic versus kinematic hardening}


\author[1]{K. J. Juul\corref{cor1}}
\ead{krjoju@mek.dtu.dk}
\cortext[cor1]{Corresponding author}

\author[2]{E. Mart\'{i}nez-Pa\~{n}eda}

\author[1]{K. L. Nielsen}

\author[1]{C. F. Niordson}

\address[1]{Department of Mechanical Engineering, Solid Mechanics, Technical University of Denmark, DK-2800 Kgs. Lyngby, Denmark}

\address[2]{Department of Engineering, Cambridge University, CB2 1PZ Cambridge, UK}

\begin{abstract}
The fracture toughness for a mode I/II crack propagating in a ductile material has been subject to numerous investigations. However, the influence of the material hardening law has received very limited attention, with isotropic hardening being the default choice if cyclic loads are absent. The present work extends the existing studies of monotonic mode I/II steady-state crack propagation with the goal to compare the predictions from an isotropic hardening model with that of a kinematic hardening model. The work is conducted through a purpose-built steady-state framework that directly delivers the steady-state solution. In order to provide a fracture criterion, a cohesive zone model is adopted and embedded at the crack tip in the steady-state framework, while a control algorithm for the far-field, that significantly reduces the number of equilibrium iterations is employed to couple the far-field loading to the correct crack tip opening. Results show that the steady-state fracture toughness (shielding ratio) obtained for a kinematic hardening material is larger than for the corresponding isotropic hardening case. The difference between the isotropic and kinematic model is tied to the non-proportional loading conditions and reverse plasticity. This also explains the vanishing difference in the shielding ratio when considering mode II crack propagation as the non-proportional loading is less pronounced and the reverse plasticity is absent.
\end{abstract}


\end{frontmatter}

\printnomenclature
\section{Introduction}
\label{S:introduction}
The influence of plastic deformation on fracture toughness has been the motivation of a large number of studies in the literature \citep[see e.g.][]{varias1993a,tvergaard1993a,Suo1993,Cleveringa2000,tvergaard2010a,Kim2012,key05,Jiang2010,Juul2016b}. The common goal has been to achieve a better understanding of the underlying mechanics that affect the toughness of ductile materials by gaining insight into the role of crack tip plasticity. Factors such as rate-dependency \citep{Landis2000}, work hardening \citep{tvergaard1992a}, strain gradients \citep{Wei1997,IJP2016}, dynamic lattice effects \citep{Freund1985}, material property mismatch \citep{cao1989a,tvergaard2002a}, or micro structure evolution \citep{Kumar2007} affect the fracture properties and determine the extent of crack propagation.

Except for the recent study of mode I cracks by \cite{JAM2018}, the majority of the published studies of crack growth under monotonic loading confine their focus to isotropic hardening materials, despite the crucial impact of the plastic material response on the shielding ratio. In fact, plastic deformation and the associated dissipation of energy is known to be the main contributor towards enhancing the fracture resistance beyond crack initiation. The far-field loading drives this process and despite being monotonic at the far boundary the conditions experienced by the material passing by the crack tip are very different. It is well documented that material entering the active plastic zone near a steadily growing mode I crack will either exit into an unloading wake or experience reverse plastic loading close to the new fracture surface. Thus, any Bauschinger effect originating from kinematic hardening must have an important influence. Though reversed plasticity does not take place in mode II crack growth, non-proportional loading for material at a distance from the crack face will be demonstrated in the results of the present study. \textcolor{black}{The present study is further motivated by the fact that kinematic hardening effects are expected to play an increasing role in modern structural materials. Composite, multiphase or refined microstructures influence the work hardening response, enhancing kinematic hardening \citep{ashby1970,formanoir2017a}}.

The main goal of the present study is to investigate how the choice of hardening model influences the fracture toughness of a steadily growing crack under monotonic mode I, mode II, and mixed mode I/II far-field conditions. \textcolor{black}{In plate tearing steadily growing cracks are encountered when the crack has propagated multiple plate thicknesses. In fact, steady-state is typically reached after crack growth on the order of seven plate thicknesses \citep{woelke2015a,Andersen_et_al_2018}. Hence, this state can dominate a significant part of the crack path in shell-like structures such as ships, air planes, and cars. To focus the effort on the part of the propagation path taking place under steady-state conditions, the framework first proposed by \cite{key01} has been adopted and extended to kinematic hardening plasticity.} The material steady-state fracture toughness, composed by the energy going into material separation as-well as energy dissipated in the surrounding material, is evaluated by introducing the cohesive traction-separation relation proposed by \cite{tvergaard1993a}. This allows for an analysis of the ratio between the external energy applied to the system and the energy specified for the fracture process (referred to as the shielding ratio). Attention is focused on the change in fracture properties when shifting from isotropic hardening to kinematic hardening. Thus, the stress evolution for material points in the vicinity of the crack tip is of particular interest as any deviation from proportional loading will be treated differently in the two types of hardening models. 

Throughout this paper, the two types of material hardening and their differences are studied for various conditions of the near tip fracture process (in terms of cohesive zone parameters). Furthermore, the origin of these differences is traced by mapping out the energy dissipation in the vicinity of the propagating crack. 
\textcolor{black}{In the present study, the material is assumed to be governed by linear hardening. This is chosen to ensure a constant tangent modulus, thus clearly bringing out the essential differences in predictions for steady-state fracture toughness between the isotropic and kinematic hardening models. }


The paper is divided into the following sections: The modified boundary value problem is presented in Section \ref{S:problem}, the material model, interface model, the algorithm controlling the far-field loading, and the numerical formulation are presented in Section \ref{S:framework}, the results are presented in Section \ref{S:results}, and lastly some concluding remarks are stated in Section \ref{S:conclusion}. Throughout, index notation, including Einstein's summation convention, is used and a superimposed dot,  $\dot{(~~)}$, denotes the time derivative.

\section{Mixed mode boundary layer problem}
\label{S:problem}
\begin{figure}[!tb]
\centering
\scalebox{1.0}{\input{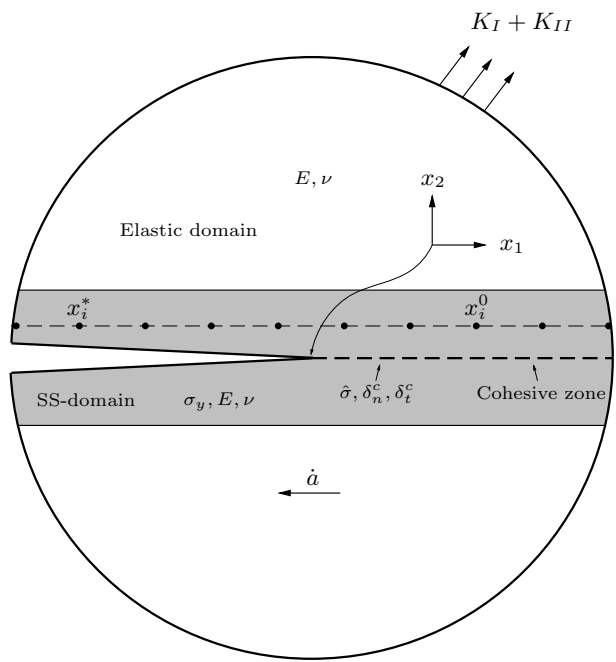}}
\caption{Mode I/II crack growth at steady-state with an embedded cohesive zone in the path of the crack.}
\label{fig:Pacman}
\end{figure}

The steady-state crack propagation study is carried out for mode I/II loading conditions under the assumption of small-scale yielding. To model the continuously growing crack, the steady-state framework presented by \cite{key01} is coupled with a cohesive zone description of fracture, employing the traction-separation law proposed by \cite{tvergaard1993a}. The problem, commonly known as the modified boundary layer problem, is modeled in a 2D plane strain setting as illustrated in Fig.~\ref{fig:Pacman} (the considered material properties are collected in Tab.~\ref{tab:mat_prop}). The domain is constructed large enough such that boundary effects do not affect the solution, and the stress intensity factors, $K_I$ and $K_{II}$, can be employed to characterize the stress-field.
The mode I/II loading condition is imposed as a far-field condition according to the elastic solution presented by \cite{Williams1957} and dictates that the stress field has the form;
\begin{align}
\sigma_{ij} = \frac{1}{\sqrt{2\pi r}}\left(K_I f_{ij}^{I}(r,\theta)+K_{II} f_{ij}^{II}(r,\theta)\right),
\label{eq:far_field}
\end{align}
where $r$ and $\theta$ are polar coordinates related to the crack tip position, $f_{ij}(r,\theta)$ are dimensionless mode functions, and $K_I$ and $K_{II}$ are the stress intensity factors representing the mode I and II contributions, respectively. Throughout this work, it is assumed that the crack propagates in a straight line (along the $x_1$-direction in Fig.~\ref{fig:Pacman}). This assumption, common to other mixed-mode crack propagation analyses \citep[see e.g.][]{tvergaard2010a}, constitutes an approximation under mode II dominated loading conditions. 

The steady-state fracture toughness is quantified by the so-called crack tip shielding ratio, $K_{ss}/K_0$, which is the stress intensity factor for steady-state crack growth, $K_{ss}$, normalized by the stress intensity factor for crack initiation, $K_{0}$. The reference stress intensity factor, $K_0$, is defined as;
\begin{align}
    K_0 = \sqrt{\frac{E\Gamma_{0}}{1-\nu^2}}
\end{align}
where $E$ is Young's modulus, $\nu$ is Poisson's ratio, and $\Gamma_0$ is the fracture energy (work of separation of the cohesive zone model). Moreover, any length quantity in the present study is normalized by the reference plastic zone size, $R_0$;
\begin{align}
R_0 &= \frac{1}{3\pi}\left(\frac{K_0}{\sigma_y}\right)^2
\label{eq:R_0}
\end{align}
where $\sigma_y$ is the initial yield stress of the material.
\begin{table}[tb!]
  \centering
    \begin{tabular}{cll}
    \toprule
    Parameter & Significance & Value \\
    \midrule
    $\sigma_y/E$  & Yield strain & 0.003 \\
    $\hat{\sigma}$  & Peak normal traction & $0.3-5.3\sigma_y$ \\
    $\nu$     & Poisson's ratio & 0.33 \\
    $E/E_t$     & Tangent modulus & $10-100$ \\
    $\lambda_1$  & Shape parameter & $0.15$ \\
    $\lambda_2$  & Shape parameter & $0.5$ \\
    \bottomrule
    \end{tabular}%
\caption{Material Properties.}  
  \label{tab:mat_prop}%
\end{table}

\section{Constitutive relations and modeling}
\label{S:framework}

\subsection{Traction-separation relation}
\label{sec:trac_sep}
\begin{figure}[!tb]
\centering
\scalebox{2.0}{\input{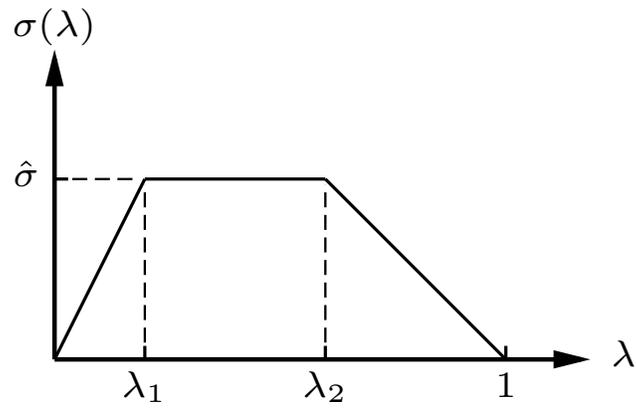}}
\caption{Traction separation relation governing the cohesive zone.}
\label{fig:Trac_law}
\end{figure}

The traction-separation relation employed is adopted from \cite{tvergaard1992a,tvergaard1993a}. Accordingly, the traction energy potential is defined as;
\begin{align}
    \Phi(\delta_t,\delta_n) = \delta_n^c \int_0^\lambda \sigma(\lambda')\text{d}\lambda'
\end{align}
where $\sigma(\lambda)$ is the traction shown in Fig.~\ref{fig:Trac_law} as a function of the non-dimensional measure of separation, $\lambda$. The non-dimensional crack separation is defined as; $\lambda = \sqrt{(\delta_n/\delta_n^c)^2+(\delta_t/\delta_t^c)^2}$, where $\delta_n$ and $\delta_t$ denote the actual separation in the normal and tangential directions, respectively, and the quantities with superscript $c$ denote the corresponding critical values. Thus, as illustrated in Fig.~\ref{fig:Trac_law}, the bond between two nodes completely breaks at $\lambda=1$.
From the traction energy potential, the normal and tangential tractions in the fracture process zone are given as;
\begin{align}
    T_n = \frac{\partial\Phi}{\partial\delta_n} = \frac{\sigma(\lambda)}{\lambda}\frac{\delta_n}{\delta_n^c} \quad \text{and}\quad T_t = \frac{\partial\Phi}{\partial\delta_t} = \frac{\sigma(\lambda)}{\lambda}\frac{\delta_n^c}{\delta_t^c}\frac{\delta_t}{\delta_t^c}.
\end{align}
Finally, the work of separation per unit area of interface (the fracture energy) is defined as;
\begin{align}
    \Gamma_0 = \frac{1}{2}\hat{\sigma}\delta_n^c(1-\lambda_1+\lambda_2)
\end{align}
where $\hat{\sigma}$ denotes the peak traction (cohesive strength) shown in Fig.~\ref{fig:Trac_law}.

\subsection{Constitutive models}
The framework relies on an infinitesimal strain formulation (both for the isotropic and kinematic model). The small strain formulation has been chosen because previous finite strains studies have shown that the crack propagates at relatively small deformations \citep{tvergaard1992a} for the selected values of the cohesive strength. This is also seen in the work by e.g., \cite{Wei1997,JAM2018} where finite strain results are precisely reproduce by a small strain framework. In the infinitesimal strain formulation, the total strains, $\varepsilon_{ij}$, are determined from the displacement gradients; $\varepsilon_{ij} = (u_{i,j}+u_{j,i})/2$. The total strains consist of an elastic part, $\varepsilon_{ij}^e$, and plastic part, $\varepsilon_{ij}^p$, which for an additive split gives the following relationship; $\varepsilon_{ij} = \varepsilon_{ij}^e + \varepsilon_{ij}^p$.
Subsequently, the stress field in the rate-independent model is determined from the tensor of instantaneous moduli, $L_{ijkl}$, and the total strain as;
\begin{equation}
    \dot{\sigma}_{ij} = L_{ijkl}\dot{\varepsilon}_{kl}.
    \label{eq:stress_rela}
\end{equation} 

Throughout this work, the material behavior is assumed to be governed by linear hardening such that the tangent modulus, $E_t$, remains constant and given as a fraction of Young's modulus, $E$. The tangent modulus, $E_t$, enters the instantaneous moduli, $L_{ijkl}$, in Eq.~\eqref{eq:Lijkl}.


\subsubsection{Isotropic hardening}
\begin{figure}[!tb]
\centering
\scalebox{1.0}{\input{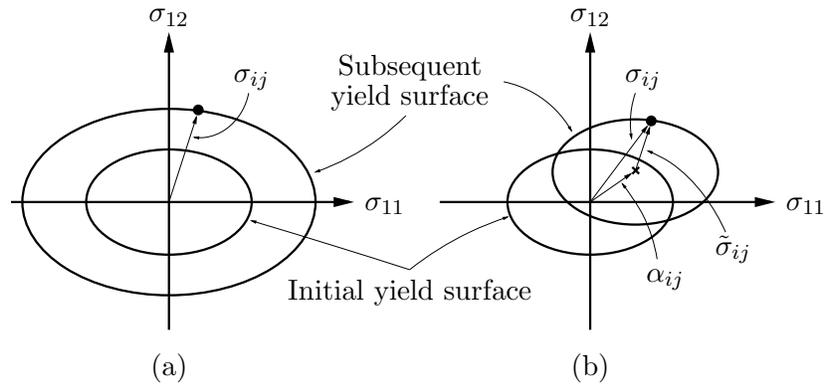}}
\caption{Evolution of yield surface for (a) isotropic hardening and (b) kinematic hardening \citep{tvergaardbook}.}
\label{fig:Yield_surface}
\end{figure}

The isotropic model does not consider the Bauschinger effect, as the yield surface expands isotropically in all directions (see Fig.~\ref{fig:Yield_surface}a) while maintaining its origin in stress space. The von Mises yield criterion, employed in the present study, takes the form
\begin{align}
    F(\sigma_{ij}) = \frac{3}{2}s_{ij}s_{ij} - (\sigma_e)^2_{\text{max}} = 0,
\end{align}
where $s_{ij}$ is the deviatoric stress and $\sigma_e$ is the von Mises stress. 
In the context of an incremental formulation, the active plastic zone can be evaluated by integrating the total stress in time (Eq.~\eqref{eq:stress_rela}), followed by an evaluation of the criterion for plasticity;
\[
    \beta= 
\begin{dcases}
    1,& \text{for } \sigma_{e} = (\sigma_{e})_{\text{max}}  \text{  and  } \dot{\sigma}_{e} \ge 0\\
    0,& \text{for } \sigma_{e} < (\sigma_{e})_{\text{max}}  \text{  or  } \dot{\sigma}_{e} < 0
\end{dcases}
\]
where $\beta=1$ indicates a material point governed by plastic loading and $\beta = 0$ indicates a material point governed by elastic unloading (applied in Eq.~\eqref{eq:Lijkl}).

\subsubsection{Kinematic hardening}
%
For kinematic hardening, where the yield surface translates in stress space (see Fig.~\ref{fig:Yield_surface}b), the yield condition reads;
\begin{align}
    F(\sigma_{ij},\alpha_{ij}) = \frac{3}{2}\tilde{s}_{ij}\tilde{s}_{ij} - (\sigma_y)^2 = 0,
    \label{eq:kin_F}
\end{align}
where symbols with ( $\tilde{ }$ ) denote stress quantities related to the local origin of the translating yield surface, representing the well-known Bauschinger effect. 
The origin of the translating yield surface, tracked through the back stress, $\alpha_{ij}$, is used to establish the local stress;
\begin{equation}
  \tilde{\sigma}_{ij}=\sigma_{ij}-\alpha_{ij}  
\end{equation}
in the yield function, Eq.~\eqref{eq:kin_F}, through the deviatoric stress, $\tilde{s}_{ij}$. The translation of the yield surface is modeled through an evolution law. Several evolution laws exist in the literature, ranging from simple models to very sophisticated models including effects such as ratchetting and shakedown \citep{Lemaitre1990}. For fracture problems under monotonic loading, reversed loading is expected in the wake of the leading active plastic zone, but repeated cyclic loading is not taking place which means that ratchetting and shakedown are not relevant to the present study. Consequently, the well-known evolution law by \cite{ziegler1959a} is chosen, as it includes all the necessary features. Thus, the back stress evolves as;
\begin{align}
    \dot{\alpha}_{ij} = (\sigma_{ij}-\alpha_{ij})\dot{\mu}
\end{align}
where $\dot{\mu}$ is a proportionality coefficient which is defined as;
\begin{equation}
  \dot{\mu}=\frac{3}{2} \frac{\dot{\sigma}_{ij}\tilde{s}_{ij}}{\sigma_y^2}  
\end{equation}
in the present study. For kinematic hardening, where the yield surface translates but maintain its size, the criterion for plasticity is therefore evaluated according to;
\[
    \beta= 
\begin{dcases}
    1,& \text{for } \tilde{\sigma}_{e} = \sigma_{y}  \text{  and  } \tilde{s}_{kl}\dot{\sigma}_{kl} \ge 0\\
    0,& \text{for } \tilde{\sigma}_{e} < \sigma_{y}  \text{  or  } \tilde{s}_{kl}\dot{\sigma}_{kl} < 0
\end{dcases},
\]

\subsection{Steady-state framework}
The fracture toughness at steady-state, $K_{ss}$, can be approximated with traditional incremental numerical methods by computing the crack growth resistance curve.
However, such methods often suffer from convergence issues and are inefficient as they are forced through the transient regime of a problem before reaching the steady-state. To avoid such issues, a steady-state framework is employed, leading to accurate predictions of the steady-state fracture toughness at a fraction of the computational cost. The steady-state framework presented builds upon an extension of the procedure suggested by \cite{key01} to account for a kinematic hardening law. 

The pivotal step is to utilize the nature of a steady-state problem to determine the history dependent field quantities. The steady-state condition is noticed for an observer located at the tip of a continuously growing crack when the field quantities that surrounds the crack tip are no longer subject to changes. Any time derived quantity, $\dot{f}$, in the constitutive equations are then transformed into spatial derivatives, through the crack propagation speed, $\dot{a}$, in the direction of the material flow (negative $x_1$-direction, see Fig.~\ref{fig:Pacman}) according to the relation;
\begin{equation}
    \dot{f} = -\dot{a} \frac{\partial f }{\partial x_1}.
\end{equation}
Thus, any total quantity at a given material point ($x_1^*,x_2^*$), is evaluated through spatial integration, starting upstream in the undeformed elastic material ahead of the crack tip ($x_1^0,x_2^*$), and following a streamline (material flow line) until it reaches the point of interest ($x_1^*,x_2^*$) downstream \citep[see e.g.][]{Juul2016b,Juul2016a}. Thus, the loading history at a given material point ($x_1^*,x_2^*$) is retrieved from all the upstream points along the streamline, representing earlier states.

The steady-state framework is based on the conventional principle of virtual work (PWV) for quasi-static problems;
\begin{equation}
\int_V \mathscr{L}_{ijkl}\varepsilon_{kl}\delta\varepsilon_{ij}\text{d}V + \int_{S_c} T_i\delta u_i \text{d}S_c = \int_S t_i\delta u_i \text{d}S + \int_V \mathscr{L}_{ijkl}\varepsilon_{kl}^p\delta\varepsilon_{ij}\text{d}V
\label{eq:VWP_disp}
\end{equation}
where $t_i = \sigma_{ij} n_j$ is the surface traction, $T_i$ is the traction from the traction-separation law, and $\mathscr{L}_{ijkl}$ is the isotropic elastic stiffness tensor. The volume analyzed is denoted $V$, $S_c$ is the interface (cohesive) surface, and $S$ is the bounding surface, with $n_j$ denoting the unit outward normal vector.

The algorithm employed for the spatial integration procedure is outlined below, where it is seen that the main difference is in the integration of the back stress in step 4 ($m$ refers to the iteration number):
\begin{enumerate}
\item Use the plastic strains, $\varepsilon_{ij}^{p(m-1)}$, to determine the current displacement field, $u_{i}^{(m)}$ ($\varepsilon_{ij}^{p(m-1)}$ is assumed to be zero in the first iteration).
\item Determine the total strain, $\varepsilon_{ij}^{(m)}$, from the displacement field, $u_{i}^{(m)}$.
\item Determine the total stress field outside the streamline domain:
\begin{enumerate}[(i)]
    \item Stresses can be determined directly from the total strain ($\varepsilon_{ij}^{p(m)} = 0$):
    \begin{align}
        \sigma_{ij}^{(m)} = \mathscr{L}_{ijkl}\varepsilon_{kl}^{(m)}
    \end{align}
    where $\mathscr{L}_{ijkl}$ is the elastic stiffness tensor. 
    \end{enumerate}
\item Determine the stresses inside the streamline domain:
\begin{enumerate}[(i)]
    \item Determine the spatial derivative of the stress: 
    \begin{align}
    \frac{\partial \sigma_{ij}^{(m)}}{\partial x_1} = L_{ijkl}^{(m)}\frac{\partial \varepsilon_{kl}^{(m)}}{\partial x_1}
    \end{align}
    where $L_{ijkl}$ is the tensor of instantaneous moduli. 
    \item Determine the spatial derivative of the back stress (only for kinematic hardening):
    \begin{align}
        \frac{\partial \alpha_{ij}^{(m)}}{\partial x_1} = (\sigma_{ij}-\alpha_{ij})\frac{\partial \mu^{(m)}}{\partial x_1} \quad \text{with} \quad \frac{\partial \mu^{(m)}}{\partial x_1} = \frac{3}{2}\frac{\tilde{s}_{ij}}{\sigma_y^2}\frac{\partial \sigma_{ij}^{(m)}}{\partial x_1}
    \end{align} 
\item Perform spatial integration along streamlines:
    \begin{align}
        \sigma_{ij}^{(m)} = \int_{x_1^0}^{x_1^*} \frac{\partial \sigma_{ij}^{(m)}}{\partial x_1} \text{d}x_1 \quad \text{and} \quad \alpha_{ij}^{(m)} = \int_{x_1^0}^{x_1^*} \frac{\partial \alpha_{ij}^{(m)}}{\partial x_1} \text{d}x_1
    \end{align}
\end{enumerate}
\item Determine the plastic strain field, $\varepsilon_{ij}^{p(m)} = \varepsilon_{ij}^{(m)} - \mathscr{M}_{ijkl}\sigma_{kl}^{(m)}$, in the streamline domain, with $\mathscr{M}_{ijkl}$ being the elastic compliance tensor. 
\item Repeat 1 to 5 until solution is converged.
\end{enumerate}

In the algorithm, the following constitutive tensors have been applied which include the elastic stiffness tensor,
\begin{align}
\mathscr{L}_{ijkl} &= \frac{E}{1+\nu}\left[\frac{1}{2}(\delta_{ik}\delta_{jl}+\delta_{il}\delta_{jk})+\frac{\nu}{1-2\nu}\delta_{ij}\delta_{kl}\right],
\end{align}
the elastic compliance tensor,
\begin{align}
\mathscr{M}_{ijkl} &= \frac{1}{E}\left[\frac{1+\nu}{2}(\delta_{ik}\delta_{jl}+\delta_{il}\delta_{jk})-\nu \delta_{ij}\delta_{kl}\right],
\end{align}
and the isotropic tensor of instantaneous moduli,
\begin{align}
L_{ijkl} = \mathscr{L}_{ijkl} - \beta\frac{3}{2}\frac{E/E_t-1}{E/E_t-(1-2\nu)/3}\frac{s_{ij}s_{kl}}{\sigma_e^2},
\label{eq:Lijkl}
\end{align}
where it should be noted that in the tensor of instantaneous moduli, $s_{ij}s_{kl}/\sigma_e^2$ is substituted with $\tilde{s}_{ij}\tilde{s}_{kl}/\sigma_y^2$ for the kinematic hardening model.

The numerical stability of the steady-state algorithm is in general better than for incremental frameworks, although, certain problems can arise in areas with steep gradients such as at the crack tip. In order to limit such stability issue of the algorithm a minor change has been made to the original procedure by \cite{key01} following the suggestion by \cite{Niordson2001} and \cite{key05}, where a sub-increment procedure between Gauss points has been introduced in the spatial integration scheme.

\subsection{Cohesive elements and traction-separation relation}
\label{subsec:cohesive}
The implementation of the cohesive zone builds upon the standard case presented by e.g. \cite{busto2017a}. However, for the steady-state framework, minor modifications are introduced when building both the stiffness matrix and the right-hand side of the equation system. When discretizing the virtual work principle in Eq. \eqref{eq:VWP_disp}, the nonlinear part of the contribution from the cohesive elements is moved to the right-hand side and acts as a force term. Hence, the discretized system reads;
\begin{align}
\begin{split}
    \left(\int_V [B]^T [\mathscr{L}] [B] \text{d}V + \int_{S_c} [B_c]^T \frac{\partial \{T^{\text{ini}}\}}{\partial \{\delta\}} [B_c] \text{d}S_c \right) & \{U\} \\ = \int_{S} [N]^T\{t\} \text{d}S + \int_V [B]^T [\mathscr{L}] \{\varepsilon^p\} \text{d}V + \int_V [B_c]^T &(\{T^{\text{lin}}\}-\{T^{\text{act}}\}) \text{d}S_c
\end{split}
\end{align}
where $\partial \{T^{\text{ini}}\}/\partial \{\delta\}$ is the initial slope of the traction-separation curve in the region $\lambda <\lambda_1$ (see Fig.~\ref{fig:Trac_law}), $T^{\text{lin}}_{i}$ is the corresponding traction predicted if a linear curve with the initial slope is followed, and $T^{\text{act}}_{i}$ is the traction obtained from the actual curve in Fig.~\ref{fig:Trac_law}. The standard strain-displacement and shape function matrices, $[B{]}$, $[B_c{]}$ and $[N{]}$, are given in the work by \cite{busto2017a}.
The partition of the cohesive contribution is necessary for the steady-state framework because the equations are no longer expressed on incremental form. However, the partition also entails a considerable reduction in the computational cost as the system matrix only needs to be built and factorized once, with subsequent iterations relying solely on back-substitution.


In the present study, the equation system is discretized using quadratic 8-node isoparametric elements evaluated through $2\times 2$ Gauss points and quadratic 6-node isoparametric cohesive elements evaluated through 8 Gauss points.

\subsection{Control algorithm for the boundary layer problem}
\label{S:control}
The far-field required to drive crack propagation in an elastic-plastic solid is generally unknown and, thus, to efficiently study the steady-state fracture toughness (the shielding ratio) it is necessary to implement a scheme to control the applied far-field loading such that the energy needed for steady-state crack propagation is provided. For this purpose, several techniques have been proposed in the literature. 
In the present study, the idea employed by \cite{Segurado} and \cite{Paneda2017a} is adopted as it offers fast and stable convergence while at the same time lending itself nicely to implementation in the developed numerical scheme. The technique is here generalized to treat problems with mode mixity. The key idea behind the procedure is to link the loading history to a monotonically increasing parameter that is not affected by potential instabilities. For the present investigation of steady-state crack growth such a parameter could be the crack tip opening displacements (normal and tangential separation) as such quantities should be non-decreasing at all times during loading - or, more precisely, kept fixed in the steady-state framework \citep[schemes with other or more parameters included can also be found in][]{Segurado}. 

\begin{figure}[!tb]
\centering
\scalebox{1}{\input{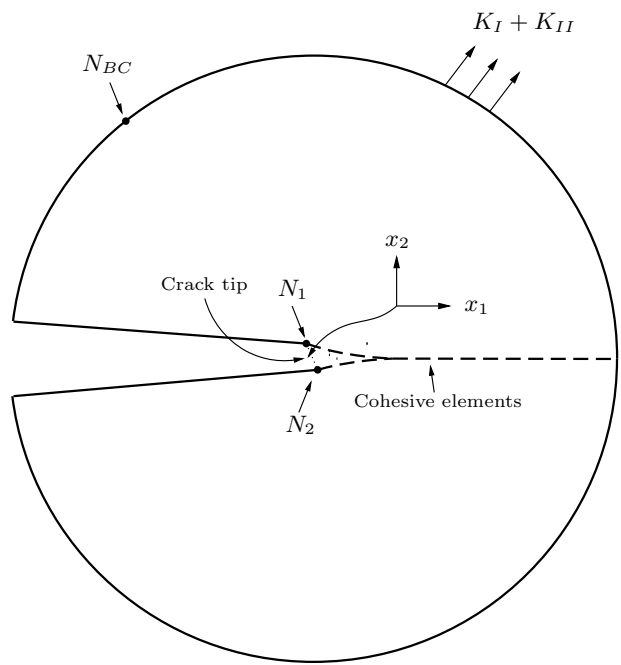}}
\caption{Illustration of linked nodes in the control algorithm.}
\label{fig:Control}
\end{figure}

Once suitable monotonically increasing parameters have been chosen, a connection between these parameters and the load at the outer far boundary must be established. The connection is introduced through; i) two constraint equations (see Eq.~\eqref{eq:const_1}) that ensure the prescribed crack tip opening, ii) a global equilibrium equation which ensures that the loading on the crack tip, enforced by the constraint equations, is balanced out by the loading on the outer far boundary, and iii) a set of geometrical constraints that ensure a smooth distribution of the far-field loading according to the elastic solution by~\cite{Williams1957}. The two additional constraint equations are introduced by adding two additional degrees of freedom, $Q_n$ and $Q_t$, to the system (with corresponding right-hand sides, $\Delta_n$ and $\Delta_t$). The desired crack tip opening, $\Delta_n$ and $\Delta_t$, is then enforced through the following displacement constraints;
\begin{align}
    u_{1}^{(N_1)}-u_{1}^{(N_2)} = \Delta_t \quad \text{and}\quad u_{2}^{(N_2)}-u_{2}^{(N_2)} = \Delta_n,
    \label{eq:const_1}
\end{align}
where $N_1$ and $N_2$ refers to the two nodes located at the crack tip on each side of the crack plane (see Fig.~\ref{fig:Control}). 
In the case of a pure mode I loaded crack, the enforced tip displacements are; $\Delta_n=\delta_n^c$ and $\Delta_t=0$, while the pure mode II crack is analyzed by enforcing; $\Delta_n=0$ and $\Delta_t=\delta_t^c$. Here, $\delta_n^c$ and $\delta_t^c$ are the critical normal and tangential separation, respectively, related to the cohesive traction-separation relation (see Section~\ref{subsec:cohesive}).

The constraint equations in Eq.~\eqref{eq:const_1} introduce reaction forces at the crack tip that drive the opening. This is, however, artificial and does not resemble crack growth under far-field mode I/II loading. To ensure proper far-field loading of the crack tip, a coupling to the outer far boundary is created through two global equilibrium considerations such that the reaction force at the crack tip becomes zero. This is enforced by adding the following contributions to the system matrix, $K$, for a given node $m$;
\begin{align}
	K{{((N_{BC})_1,m)}}u^{(m)}+Q_t&=0 \\
	K{{((N_{BC})_2,m)}}u^{(m)}+Q_n&=0
\end{align}
where $(N_{BC})_i$ refers to the global degrees of freedom of an arbitrary node, $N_{BC}$, on the outer far boundary (see Fig.~\ref{fig:Control}). 
%
%
The coupling ensures that the chosen node on the far boundary, $N_{BC}$, will displace according to the prescribed crack tip opening, as dictated by equilibrium, i.e. the far-field becomes an outcome of the equilibrium solution. Finally, a set of geometric constraints are defined to ensure that the displacement of all the nodes at the outer boundary is consistent with the displacements of the arbitrary node, $N_{BC}$. The geometric constraint is determined from the elastic far-field solution to the boundary layer problem presented in Section \ref{S:problem}. In terms of the displacement field, the elastic far-field of node $m$, at the outer far boundary, reads;
\begin{align}
\begin{aligned}
\label{eq:ux}
u_{1}^{(m)}(r_{(m)}, \theta_{(m)}) &= C_{(m)} (K_{I}f_{K_I}^{u_{1}^{(m)}}(r_{(m)}, \theta_{(m)}) + K_{II} f_{K_{II}}^{u_{1}^{(m)}}(r_{(m)}, \theta_{(m)})) \\ 
u_{2}^{(m)}(r_{(m)}, \theta_{(m)}) &= C_{(m)} (K_{I}f_{K_I}^{u_{2}^{(m)}}(r_{(m)}, \theta_{(m)}) - K_{II} f_{K_{II}}^{u_{2}^{(m)}}(r_{(m)}, \theta_{(m)}))
\end{aligned}
\end{align}
where $C_{m} = 1/(2\mu)\sqrt{r_{(m)}/(2\pi)}$ and $f$ are displacement mode functions. Thus, the stress intensity factors, $K_I$ and $K_{II}$, corresponding to a known displacement, $u_i^{(m)}$, of a given node, $m$, on the outer far boundary can be determined by inverting Eq.~\eqref{eq:ux}, such that;
\begin{align}
\begin{aligned}
    K_{I} &= \frac{f_{K_{II}}^{u^{(m)}_{2}}u^{(m)}_{1} + f_{K_{II}}^{u^{(m)}_{1}}u^{(m)}_{2}}{C_{(m)}(f_{K_{II}}^{u^{(m)}_{1}}f_{K_{I}}^{u^{(m)}_{2}}+f_{K_{II}}^{u^{(m)}_{2}}f_{K_{I}}^{u^{(m)}_{1}})} \\
    K_{II} &= - \frac{f_{K_{I}}^{u^{(m)}_{2}}u^{(m)}_{1} - f_{K_{I}}^{u^{(m)}_{1}}u^{(m)}_{2}}{C_{(m)}(f_{K_{II}}^{u^{(m)}_{1}}f_{K_{I}}^{u^{(m)}_{2}}+f_{K_{II}}^{u^{(m)}_{2}}f_{K_{I}}^{u^{(m)}_{1}})}
\end{aligned}
\end{align}
That is, the $K$-field applied to the outer far boundary can be linked to the displacement of the arbitrary node, $N_{BC}$, and subsequently to the remaining nodes on the outer far boundary. In this way, the multi-point geometric constraint is stated as;
\begin{align}
\begin{aligned}
%
%
\begin{split}
    u^{(m)}_{1} &- \frac{C_{(m)}}{C_{(N_{BC})}} \frac{f_{K_{II}}^{u^{(m)}_{1}}f_{K_{I}}^{u^{(N_{BC})}_{2}}+f_{K_{I}}^{u^{(m)}_{1}}f_{K_{II}}^{u^{(N_{BC})}_{2}}}{f_{K_{II}}^{u^{(N_{BC})}_{1}}f_{K_{I}}^{u^{(N_{BC})}_{2}}+f_{K_{II}}^{u^{(N_{BC})}_{2}}f_{K_{I}}^{u^{(N_{BC})}_{1}}}u_{1}^{(N_{BC})} \\ 
    &- \frac{C_{(m)}}{C_{(N_{BC})}} \frac{f_{K_{I}}^{u^{(m)}_{1}}f_{K_{II}}^{u^{(N_{BC})}_{1}}-f_{K_{II}}^{u^{(m)}_{1}}f_{K_{I}}^{u^{(N_{BC})}_{1}}}{f_{K_{II}}^{u^{(N_{BC})}_{1}}f_{K_{I}}^{u^{(N_{BC})}_{2}}+f_{K_{II}}^{u^{(N_{BC})}_{2}}f_{K_{I}}^{u^{(N_{BC})}_{1}}}u_{2}^{(N_{BC})}= 0 
\end{split} \\
\begin{split}
    u^{(m)}_{2} &- \frac{C_{(m)}}{C_{(N_{BC})}} \frac{f_{K_{I}}^{u^{(m)}_{2}}f_{K_{II}}^{u^{(N_{BC})}_{2}}-f_{K_{II}}^{u^{(m)}_{2}}f_{K_{I}}^{u^{(N_{BC})}_{2}}}{f_{K_{II}}^{u^{(N_{BC})}_{1}}f_{K_{I}}^{u^{(N_{BC})}_{2}}+f_{K_{II}}^{u^{(N_{BC})}_{2}}f_{K_{I}}^{u^{(N_{BC})}_{1}}}u_{1}^{(N_{BC})} \\
    &- \frac{C_{(m)}}{C_{(N_{BC})}} \frac{f_{K_{I}}^{u^{(m)}_{2}}f_{K_{II}}^{u^{(N_{BC})}_{1}}+f_{K_{II}}^{u^{(m)}_{2}}f_{K_{I}}^{u^{(N_{BC})}_{1}}}{f_{K_{II}}^{u^{(N_{BC})}_{1}}f_{K_{I}}^{u^{(N_{BC})}_{2}}+f_{K_{II}}^{u^{(N_{BC})}_{2}}f_{K_{I}}^{u^{(N_{BC})}_{1}}}u_{2}^{(N_{BC})}= 0
\end{split}
\end{aligned}
\end{align}

The multi-point geometric constraint is enforced directly on the stiffness matrix. In combination with the constraint equation in Eq.~\eqref{eq:const_1} this ensures a distributed loading on the outer far boundary according to the $K$-field required to enforce the prescribed crack tip displacements ($\Delta_n$ and $\Delta_t$).


\section{Results}
\label{S:results}
Two types of analyses are conducted to gain better insight into how the hardening model affects the fracture toughness at steady-state crack growth in an elastic-plastic solid. First, the shielding ratio, $K_{ss}/K_0$, for various fracture process zone conditions (controlled by the cohesive zone) is investigated to quantify the differences between isotropic and kinematic hardening predictions (Section \ref{Sec:ResultsShieldingRatio}). Results are obtained for pure mode I, pure mode II and mixed-mode steady-state crack propagation. 
Secondly, the origin of the notable differences observed in the shielding ratio is investigated through the energy dissipation density in the isotropic and kinematic hardening materials (Section \ref{Sec:ResultsPlasticZone}). The energy dissipation density is investigated in an attempt to identify which regions of the active plastic zone in the vicinity of the crack tip that primarily controls shielding by adding to the steady-state fracture toughness. This investigation is conducted by tracing material points traveling along the crack path (in the $x_1$-direction) at different distances from the crack face.

The study employs a mesh with a total of 310,000 elements in the entire domain, where approximately 60,000 of the elements are located in the region of the main plastic zone. Furthermore, the critical normal and tangential separation will be related to the mesh such that; $\delta^c_n = \delta^c_t = 0.2L_{e,\text{min}}$, where $L_{e,\text{min}}$ is the minimum element length in the domain.

\subsection{Crack tip shielding ratio}
\label{Sec:ResultsShieldingRatio}
To quantify the differences between isotropic and kinematic hardening, the shielding ratio, $K_{ss}/K_{0}$, is studied by employing the cohesive zone model presented in Section~\ref{sec:trac_sep}. The cohesive zone model ensures that the energy release rate required for crack propagation is identical for both the isotropic and kinematic material while the far-field is scaled accordingly. 

The crack tip shielding ratio, $K_{ss}/K_{0}$, is presented as a function of the normalized peak traction, $\hat{\sigma}/\sigma_y$, in Fig.~\ref{fig:shield_I} for a pure mode I crack ($K_{II}=0$) for both isotropic (dashed lines) and kinematic hardening (solid lines). 
For low hardening ($E_t=E/100$) it is seen that the shielding ratio is almost identical for the two materials as hardening has limited influence on the predicted stress level (close to perfectly plastic). That is, the yield surface remains close to identical for the two types of material hardening. However, a significant shift is seen for the isotropic model when the strain hardening is increased to $E_t=E/20$, while the effect in the kinematic model is small. Similarly, by choosing an even higher hardening, $E_t=E/10$, an even greater shift is seen for the isotropic material, whereas the shift for the kinematic material is much less pronounced. It is important to notice that for fixed strain hardening and peak traction in the cohesive zone, the kinematic material always predicts either an equal or greater shielding ratio compared to that of the isotropic material model. Furthermore, it is seen that shield ratio for the kinematic hardening solid appears to be unbounded for peak tractions larger than 2.9, 3.2 and 4.4 times the yield stress for $E_t=E/100$, $E_t=E/20$ and $E_t=E/10$, respectively.
\begin{figure}
\centering
\begin{tikzpicture}
    \begin{axis}[
        legend style={font=\tiny},
        legend pos=outer north east,
        grid,
         xmin= 1.8,   xmax=5.5,
	    ymin=0.8,   ymax=3,
	    mark repeat=2,
	   legend cell align={left},
        xlabel=$\hat{\sigma}/\sigma_y$,
        ylabel=$K_{ss}/K_0$]

     \addplot [line width=0.5mm,smooth,dashed,black,mark options={solid,fill=black},mark=triangle]  table [x expr=\thisrowno{0},y expr=\thisrowno{1}^0.5]{Iso_Et_100_I.txt};
     \addlegendentry{Iso, $E_t=E/100$}

     \addplot [line width=0.5mm,dashed,smooth,red,mark options={solid,fill=red},mark=*]  table [x expr=\thisrowno{0},y expr=\thisrowno{1}^0.5]{Iso_Et_20_I.txt};
     \addlegendentry{Iso, $E_t=E/20$}
     
     \addplot [line width=0.5mm,dashed,smooth,blue,mark options={solid,fill=blue},mark=square*]  table [x expr=\thisrowno{0},y expr=\thisrowno{1}^0.5]{Iso_Et_10_I.txt};
     \addlegendentry{Iso, $E_t=E/10$}
     
     \addplot [line width=0.5mm,smooth,black,mark options={solid,fill=black},mark=triangle]  table [x expr=\thisrowno{0},y expr=\thisrowno{1}^0.5]{Kin_Et_100_I.txt};
     \addlegendentry{Kin, $E_t=E/100$}     
     
	 \addplot [line width=0.5mm,smooth,red,mark options={solid,fill=red},mark=*]  table [x expr=\thisrowno{0},y expr=\thisrowno{1}^0.5]{Kin_Et_20_I.txt};
     \addlegendentry{Kin, $E_t=E/20$}     
          
     \addplot [line width=0.5mm,smooth,blue,mark options={solid,fill=blue},mark=square*]  table [x expr=\thisrowno{0},y expr=\thisrowno{1}^0.5]{Kin_Et_10_I.txt};
     \addlegendentry{Kin, $E_t=E/10$}

    \end{axis}
    \end{tikzpicture}
\caption{Shielding ratio for a mode I loaded crack ($K_{II} = 0$) propagating in isotropic and kinematic hardening materials, respectively, at various peak tractions, $\hat{\sigma}$, and tangent modulus, $E_t$.} 
\label{fig:shield_I}    
\end{figure}

The shielding ratio for a pure mode II loaded crack ($K_{I}=0$) is presented in Fig.~\ref{fig:shield_II}. Contrary to the pure mode I crack, a much less significant effect is observed when changing from an isotropic to a kinematic hardening model. As expected, the curves are almost identical for the case of $E_t=E/100$ (and essentially coincides in the figure), but the difference in the predicted shielding ratio remains limited when increasing the strain hardening. When compared to the mode I study, it is also observed that similar values of the steady-state fracture toughness are attained for significantly lower cohesive strengths. The reason for this is that the stress triaxiality close to the crack tip is significantly smaller in mode II, promoting plastic deformation at lower load levels, as specified in terms of the stress intensity factor. Furthermore, as discussed by \citet{tvergaard2010a}, values in closer agreement between mode I and mode II would be attained if the crack was allowed to grow along its preferred path.
%

\begin{figure}
\centering
\begin{tikzpicture}
    \begin{axis}[
        legend style={font=\tiny},
        legend pos=outer north east,
        grid,
         xmin= 0.2,   xmax=1.2,
	    ymin=0.8,   ymax=3,
	    mark repeat=2,
	   legend cell align={left},
        xlabel=$\hat{\sigma}/\sigma_y$,
        ylabel=$K_{ss}/K_0$]

     \addplot [line width=0.5mm,smooth,dashed,black,mark options={solid,fill=black},mark=triangle]  table [x expr=\thisrowno{0},y expr=\thisrowno{1}^0.5]{Iso_Et_100_II.txt};
     \addlegendentry{Iso, $E_t=E/100$}
     \addplot [line width=0.5mm,dashed,smooth,red,mark options={solid,fill=red},mark=*]  table [x expr=\thisrowno{0},y expr=\thisrowno{1}^0.5]{Iso_Et_20_II.txt};
     \addlegendentry{Iso, $E_t=E/20$}
     \addplot [line width=0.5mm,dashed,smooth,blue,mark options={solid,fill=blue},mark=square*]  table [x expr=\thisrowno{0},y expr=\thisrowno{1}^0.5]{Iso_Et_10_II.txt};
     \addlegendentry{Iso, $E_t=E/10$}

     \addplot [line width=0.5mm,smooth,black,mark options={solid,fill=black},mark=triangle]  table [x expr=\thisrowno{0},y expr=\thisrowno{1}^0.5]{Kin_Et_100_II.txt};
     \addlegendentry{Kin, $E_t=E/100$}
     \addplot [line width=0.5mm,smooth,red,mark options={solid,fill=red},mark=*]  table [x expr=\thisrowno{0},y expr=\thisrowno{1}^0.5]{Kin_Et_20_II.txt};
     \addlegendentry{Kin, $E_t=E/20$}
     \addplot [line width=0.5mm,smooth,blue,mark options={solid,fill=blue},mark=square*]  table [x expr=\thisrowno{0},y expr=\thisrowno{1}^0.5]{Kin_Et_10_II.txt};
     \addlegendentry{Kin, $E_t=E/10$}

    \end{axis}
    \end{tikzpicture}
\caption{Shielding ratio for mode II loaded crack ($K_{I} = 0$) propagating in an isotropic and a kinematic hardening material, respectively, at various peak tractions, $\hat{\sigma}$, and tangent modulus, $E_t$.} 
\label{fig:shield_II}    
\end{figure}

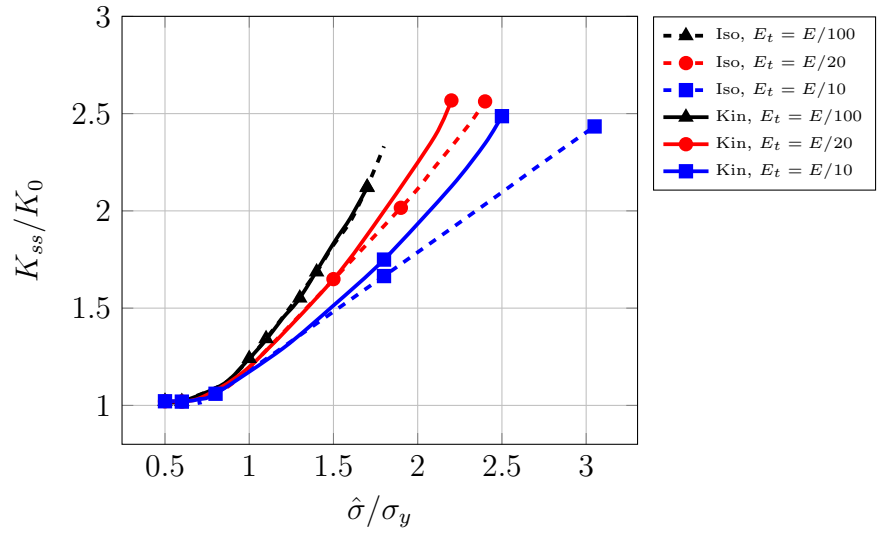
\begin{figure}
\centering
 \begin{tikzpicture}
     \begin{axis}[
         legend style={font=\tiny},
         legend pos=outer north east,
         grid,
 	    ymin=0.8,   ymax=3,
 	    mark repeat=3,
 	   legend cell align={left},
 	     xlabel=$\hat{\sigma}/\sigma_y$,
         ylabel=$K_{ss}/K_0$]

       \addplot [line width=0.5mm,mark options={solid,fill=black},mark=triangle,smooth,dashed,black]  table [x expr=\thisrowno{0},y expr=\thisrowno{1}^0.5]{Iso_Et_100_III.txt};
       \addlegendentry{Iso, $E_t=E/100$}
      \addplot [line width=0.5mm,mark options={solid,fill=red},mark=*,dashed,smooth,red]  table [x expr=\thisrowno{0},y expr=\thisrowno{1}^0.5]{Iso_Et_20_III.txt};
      \addlegendentry{Iso, $E_t=E/20$}
       \addplot [line width=0.5mm,mark options={solid,fill=blue},mark=square*,dashed,smooth,blue]  table [x expr=\thisrowno{0},y expr=\thisrowno{1}^0.5]{Iso_Et_10_III.txt};
       \addlegendentry{Iso, $E_t=E/10$}

       \addplot [line width=0.5mm,mark options={solid,fill=black},mark=triangle,smooth,black]  table [x expr=\thisrowno{0},y expr=\thisrowno{1}^0.5]{Kin_Et_100_III.txt};
       \addlegendentry{Kin, $E_t=E/100$}
       \addplot [line width=0.5mm,mark options={solid,fill=red},mark=*,smooth,red]  table [x expr=\thisrowno{0},y expr=\thisrowno{1}^0.5]{Kin_Et_20_III.txt};
       \addlegendentry{Kin, $E_t=E/20$}
       \addplot [line width=0.5mm,mark options={solid,fill=blue},mark=square*,smooth,blue]  table [x expr=\thisrowno{0},y expr=\thisrowno{1}^0.5]{Kin_Et_10_III.txt};
       \addlegendentry{Kin, $E_t=E/10$}

     \end{axis}
     \end{tikzpicture}
  \caption{Shielding ratio for mixed mode I/II loaded crack ($K_I = K_{II}$) propagating in an isotropic and kinematic hardening material, respectively, at various peak tractions and tangent modulus.} 
  \label{fig:shield_I/II}       
\end{figure}

Finally, the shielding ratio is studied for the mixed mode case where $K_I = K_{II}$. Figure~\ref{fig:shield_I/II} reveals a shielding ratio for mixed mode conditions which is located between the pure mode I and pure mode II predictions. However, it should be noted that the contribution from each mode in an elastic-plastic solid relies on the crack tip conditions and is therefore not directly reflected by the far-field mode mixity.
%

\subsection{Active plastic zones and energy dissipation}
\label{Sec:ResultsPlasticZone}
\begin{figure}[!tb]
         \centering        
 \begin{tikzpicture}
     \begin{axis}[
         legend pos=north west,
         grid,
        xmin= -5.8, 
 	    ymin=0,
 	    legend cell align={left},
         xlabel=$x_1/R_0$,
         ylabel=$x_2/R_0$]
     
      \addplot [smooth,black,dashed]  table [x expr=\thisrowno{0},y expr=\thisrowno{1}]{plastic_zone_iso_2_E.txt};
      \addlegendentry{Iso}
      \addplot [smooth,black,dashed]  table [x expr=\thisrowno{0},y expr=\thisrowno{1}]{plastic_zone_iso_3_E.txt};
      \addlegendentry{}
     
      \addplot [smooth,black]  table [x expr=\thisrowno{0},y expr=\thisrowno{1}]{plastic_zone_kin_1_E.txt};
      \addplot [smooth,black]  table [x expr=\thisrowno{0},y expr=\thisrowno{1}]{plastic_zone_kin_2_E.txt};
      \addlegendentry{Kin}
      
      \addplot [smooth,black]  table [x expr=\thisrowno{0},y expr=\thisrowno{1}]{plastic_zone_kin_3_E.txt};
     
      \legend{Iso,,Kin}
    
    
     \end{axis}
     \end{tikzpicture}
         \caption{Active plastic zones for isotropic and kinematic hardening with $E_t=E/20$ and $\hat{\sigma}/\sigma_y=3.5$ for pure mode I.} 
         \label{fig:p_zone_equal_K0}
\end{figure}
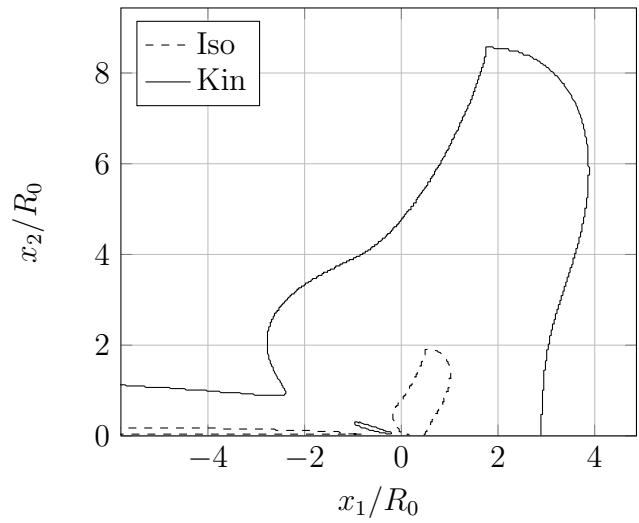

The investigation of steady-state fracture toughness showed a larger shield ratio for kinematic hardening solids, independently of the loading mode. To assess the influence on the material behavior in the vicinity of the crack tip, the plastic zone is presented in Fig.~\ref{fig:p_zone_equal_K0} for both isotropic and kinematic hardening for the mode I case with tangent modulus $E_t=E/20$ and peak traction $\hat{\sigma}/\sigma_y=3.5$. As expected, the plastic zone is significantly larger in the kinematic hardening case to accommodate the larger dissipation. 

The different energy dissipation levels in the two hardening models are investigated through the plastic zone shapes and the associated energy dissipation densities by applying identical far-field loading conditions (equal energy input to the system) in the two models. A direct consequence of this study is that the energy at the crack tip depends on the energy dissipation in each model, i.e. the fracture criterion is not identical. For this study, the cohesive elements are removed from the model i.e. the crack plane is now represented by a line of single nodes rather than dual nodes such that crack tip opening does not take place and the crack remains perfectly sharp. The level of hardening remains fixed at $E_t=E/20$. Furthermore, in the study of the plastic zones, any length quantity is normalized by Irwin's approximation of the plastic zone size $R_p$, identical to Eq.~\eqref{eq:R_0}, but with the far-field, $K$, instead of the initiation threshold $K_0$ (similarly, $\Gamma$ is computed from the far-field, $K$).

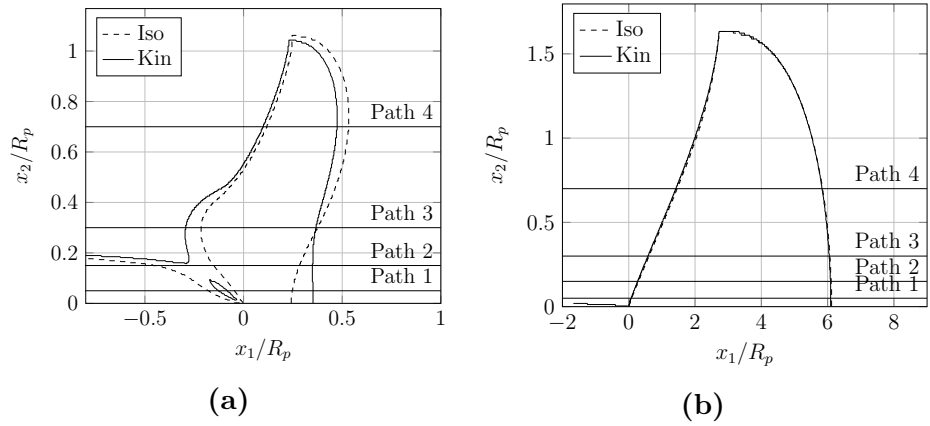
\begin{figure}[!tb]
         \centering        
         \begin{subfigure}{0.45\textwidth}
         \centering
 \resizebox{\linewidth}{!}{        
 \begin{tikzpicture}
     \begin{axis}[
         legend pos=north west,
         grid,
         xmin= -0.8,   xmax=1,
 	    ymin=0,
 	    legend cell align={left},
         xlabel=$x_1/R_p$,
         ylabel=$x_2/R_p$]
     
      \addplot [smooth,black,dashed]  table [x expr=\thisrowno{0},y expr=\thisrowno{1}]{plastic_zone_iso_1.txt};
      \addlegendentry{Iso}
      \addplot [smooth,black,dashed]  table [x expr=\thisrowno{0},y expr=\thisrowno{1}]{plastic_zone_iso_2.txt};
      \addlegendentry{}
     
      \addplot [smooth,black]  table [x expr=\thisrowno{0},y expr=\thisrowno{1}]{plastic_zone_kin_1.txt};
      \addlegendentry{Kin}
      \addplot [smooth,black]  table [x expr=\thisrowno{0},y expr=\thisrowno{1}]{plastic_zone_kin_3.txt};
     
      \draw (axis cs:-1,0.7) -- node[at end, above left]{Path 4} (axis cs:1,0.7);
      \draw (axis cs:-1,0.3) -- node[at end, above left]{Path 3} (axis cs:1,0.3);
      \draw (axis cs:-1,0.15) -- node[at end, above left]{Path 2} (axis cs:1,0.15);
      \draw (axis cs:-1,0.05) -- node[at end, above left]{Path 1} (axis cs:1,0.05);
      \legend{Iso,,Kin}
    
    
     \end{axis}
     \end{tikzpicture}
 }
              \caption{}
         \end{subfigure}
         \begin{subfigure}{0.45\textwidth}
         \centering
         \resizebox{\linewidth}{!}{
 \begin{tikzpicture}
     \begin{axis}[
         legend pos=north west,
         grid,
         xmin= -2,   xmax=9,
 	    ymin=0,
 	    legend cell align={left},
         xlabel=$x_1/R_p$,
         ylabel=$x_2/R_p$]
     
      \addplot [smooth,black,dashed]  table [x expr=\thisrowno{0},y expr=\thisrowno{1}]{plastic_zone_iso_1_II.txt};
      \addlegendentry{Iso}
     
      \addplot [smooth,black]  table [x expr=\thisrowno{0},y expr=\thisrowno{1}]{plastic_zone_kin_1_II.txt};
      \addlegendentry{Kin}
      \addplot [smooth,black]  table [x expr=\thisrowno{0},y expr=\thisrowno{1}]{plastic_zone_kin_3_II.txt};
     
      \draw (axis cs:-2,0.7) -- node[at end, above left]{Path 4} (axis cs:9,0.7);
      \draw (axis cs:-2,0.3) -- node[at end, above left]{Path 3} (axis cs:9,0.3);
      \draw (axis cs:-2,0.15) -- node[at end, above left]{Path 2} (axis cs:9,0.15);
      \draw (axis cs:-2,0.05) -- node[at end, above left]{Path 1} (axis cs:9,0.05);
      \legend{Iso,Kin}
    
    
     \end{axis}
     \end{tikzpicture}
     }
                 \caption{}
         \end{subfigure}
         \caption{Active plastic zones for isotropic and kinematic hardening in the absence of the cohesive zone with $E_t=E/20$. Here for; a) pure mode I, and b) pure mode II. The horizontal lines illustrate the paths along which the energy dissipation density has been extracted. The paths 1, 2, 3, and 4 are located at the heights; $x_2/R_p=0.05,0.15,0.3,0.7$, respectively.} 
         \label{fig:p_zone}
\end{figure}

The active plastic zones for pure mode I and mode II loading are presented in Fig.~\ref{fig:p_zone}. For a mode I crack (see Fig.~\ref{fig:p_zone}a), the effect of the kinematic model is most pronounced close to the crack tip - especially on the downstream side (left in Fig.~\ref{fig:p_zone}) of the primary active plastic zone where the material experiences reversed loading. The isotropic model is dominant, in terms of the width of the active plastic zone, further away from the crack face on the upstream side (right in Fig.~\ref{fig:p_zone}). For the mode II loaded crack (see Fig.~\ref{fig:p_zone}b), no significant differences between the plastic zone for the isotropic and kinematic hardening models are observed, consistent with the shielding ratio study. This is largely tied to the nature of a mode II loaded crack tip as reversed loading is absent and the non-proportionality is less severe compared to the mode I crack. 

To further investigate the differences between the two hardening models, the energy dissipation density, $w$, is extracted along the four horizontal paths illustrated in Fig.~\ref{fig:p_zone}. On each path, the energy dissipation density is evaluated as;
\begin{equation}
    w = -\int_{-\infty}^{\infty}\left(\sigma_{ij} \frac{\text{d}\varepsilon_{ij}^p}{\text{d}x_1}\right) \text{d}x_1,
\end{equation}
according to the streamline integration scheme. The energy dissipation density is presented in Fig.~\ref{fig:dissi_all} for mode I loading. Here, it is seen that energy dissipation density continuously grows from the onset of the plastic zone until the steady-state level is reached far behind the crack tip. Common to all paths is that the kinematic hardening entails a larger energy dissipation density. The largest difference is seen closest to the crack plane (path 1) in the unloading region immediately behind the crack tip where the kinematic model dissipates more energy than the isotropic model. The kinematic model continues to have the largest dissipation density for path 2 and 3 albeit the difference becomes progressively smaller when moving away from the crack plane as the loading becomes less complex and less severe, affecting both the level of plasticity and the non-proportionality. The trend is, however, inverted for path 4 as the isotropic solution here displays the largest dissipation density. Nevertheless, the magnitude of the energy dissipation density in this region is comparably low and has little influence on the overall dissipation.
%
\begin{figure}[!tb]
         \centering        
         \begin{subfigure}{0.45\textwidth}
         \centering
 \resizebox{\linewidth}{!}{        
         \begin{tikzpicture}
     \begin{axis}[
         legend pos=north east,
         grid,
         xmin= -1,   xmax=0.7,
 	    ymin= -0.2, ymax=2,
 	    legend cell align={left},
         xlabel=$x_1/R_p$,
         ylabel=$w/(\Gamma/R_p)$]
        
      \addlegendimage{empty legend}
      \addlegendentry{\hspace{-.6cm}\textbf{\underline{Path 1}}}
   
      \addplot [line width=0.5mm,smooth,black,dashed]  table [x expr=\thisrowno{0},y expr=\thisrowno{1}/15.0796]{dissipation_iso.txt};
      \addlegendentry{Iso}
      \addplot [line width=0.5mm,stack plots=y,smooth,black]  table [x expr=\thisrowno{0},y expr=\thisrowno{1}/15.0796]{dissipation_kin.txt};
      \addlegendentry{Kin}
    
      \addplot [line width=0.5mm,stack plots=y, stack dir=minus,smooth,black,dotted]  table [x expr=\thisrowno{0},y expr=\thisrowno{1}/15.0796]{dissipation_iso.txt};
      \addlegendentry{Diff}
    
     \end{axis}
     \end{tikzpicture}
 }
         \end{subfigure}
         \begin{subfigure}{0.45\textwidth}
         \centering
         \resizebox{\linewidth}{!}{
         \begin{tikzpicture}
     \begin{axis}[
         legend pos=north east,
         grid,
         xmin= -1,   xmax=0.7,
 	    ymin= -0.2, ymax=2,
 	    legend cell align={left},
         xlabel=$x_1/R_p$,
         ylabel=$w/(\Gamma/R_p)$]
        
      \addlegendimage{empty legend}
      \addlegendentry{\hspace{-.6cm}\textbf{\underline{Path 2}}}
     
      \addplot [line width=0.5mm,smooth,black,dashed]  table [x expr=\thisrowno{0},y expr=\thisrowno{2}/15.0796]{dissipation_iso.txt};
      \addlegendentry{Iso}
      \addplot [line width=0.5mm,stack plots=y,smooth,black]  table [x expr=\thisrowno{0},y expr=\thisrowno{2}/15.0796]{dissipation_kin.txt};
      \addlegendentry{Kin}
    
      \addplot [line width=0.5mm,stack plots=y, stack dir=minus,smooth,black,dotted]  table [x expr=\thisrowno{0},y expr=\thisrowno{2}/15.0796]{dissipation_iso.txt};
      \addlegendentry{Diff}
    
     \end{axis}
     \end{tikzpicture}
     }
         \end{subfigure}
         \begin{subfigure}{0.45\textwidth}
         \centering
         \resizebox{\linewidth}{!}{
         \begin{tikzpicture}
     \begin{axis}[
         legend pos=north east,
         grid,
         xmin= -1,   xmax=0.7,
 	    ymin= -0.2, ymax=2,
 	    legend cell align={left},
         xlabel=$x_1/R_p$,
         ylabel=$w/(\Gamma/R_p)$]
     
      \addlegendimage{empty legend}
      \addlegendentry{\hspace{-.6cm}\textbf{\underline{Path 3}}}
     
      \addplot [line width=0.5mm,smooth,black,dashed]  table [x expr=\thisrowno{0},y expr=\thisrowno{3}/15.0796]{dissipation_iso.txt};
      \addlegendentry{Iso}
      \addplot [line width=0.5mm,stack plots=y,smooth,black]  table [x expr=\thisrowno{0},y expr=\thisrowno{3}/15.0796]{dissipation_kin.txt};
      \addlegendentry{Kin}
    
      \addplot [line width=0.5mm,stack plots=y, stack dir=minus,smooth,black,dotted]  table [x expr=\thisrowno{0},y expr=\thisrowno{3}/15.0796]{dissipation_iso.txt};
      \addlegendentry{Diff}
   
  \end{axis}
     \end{tikzpicture}
     }
         \end{subfigure}
         \begin{subfigure}{0.45\textwidth}
         \centering
         \resizebox{\linewidth}{!}{
         \begin{tikzpicture}
     \begin{axis}[
         legend pos=north east,
         grid,
         xmin= -1,   xmax=0.7,
 	    ymin= -0.2, ymax=2,
 	    legend cell align={left},
         xlabel=$x_1/R_p$,
         ylabel=$w/(\Gamma/R_p)$]
     
      \addlegendimage{empty legend}
      \addlegendentry{\hspace{-.6cm}\textbf{\underline{Path 4}}}
     
      \addplot [line width=0.5mm,smooth,black,dashed]  table [x expr=\thisrowno{0},y expr=\thisrowno{4}/15.0796]{dissipation_iso.txt};
      \addlegendentry{Iso}
      \addplot [line width=0.5mm,stack plots=y,smooth,black]  table [x expr=\thisrowno{0},y expr=\thisrowno{4}/15.0796]{dissipation_kin.txt};
      \addlegendentry{Kin}
    
      \addplot [line width=0.5mm,stack plots=y, stack dir=minus,smooth,black,dotted]  table [x expr=\thisrowno{0},y expr=\thisrowno{4}/15.0796]{dissipation_iso.txt};
      \addlegendentry{Diff}
    
     \end{axis}
     \end{tikzpicture}
     }
         \end{subfigure}
         \caption{Mode I energy dissipation density in the absence of the cohesive zone for isotropic and kinematic hardening, and the difference between the models, with $E_t=E/20$. } 
         \label{fig:dissi_all}
\end{figure}

To identify possible sources of the difference in the energy dissipation density for a mode I loaded crack, the translation of the kinematic yield surface is investigated. The translation of the kinematic yield surface is represented by the effective back stress;
\begin{equation}
  \alpha_e=\sqrt{3\alpha_{ij}'\alpha_{ij}'/2}.
\end{equation}
In Fig.~\ref{fig:alpha_e} it is seen that for path 1 and 2 (see Fig.~\ref{fig:p_zone}a), the yield surface continues to translate in stress space as reversed plasticity takes place behind the crack tip. Ultimately, the effective back stress profile reaches a short unloading plateau further behind the crack tip. This continued movement of the kinematic yield surface causes additional dissipation when compared to the isotropic model. As non-proportional loading or reversed plasticity takes place the kinematic hardening solid experiences a greater change to the yield surface normal (larger curvature compared to the expanded isotropic yield surface) and hence the plastic strain increment undergoes a larger change in direction (or magnitude). In fact, this observation is partially supported by \cite{tvergaard1978a}, investigating sheet metal necking using a kinematic hardening model. Here, the kinematic hardening was observed to promote localization \citep[similar to corner theories, see][]{Mear1985} due to the curvature of the kinematic yield surface. This suggests that non-proportional loading has a significant impact in a kinematic model.
For path 3 and 4, it is seen that the back stress reaches a constant value directly due to the absence of reversed plasticity in these regions. This observation, combined with less severe non-proportionality of the loading in this region, supports the smaller difference in dissipation density observed in Fig.~\ref{fig:dissi_all}. 
\begin{figure}
\centering
\begin{tikzpicture}
    \begin{axis}[
        legend style={font=\tiny},
        legend pos=north east,
        grid,
        xmin= -1,   xmax=0.7,
	    legend cell align={left},
        xlabel=$x_1/R_p$,
        ylabel=$\alpha_e/\sigma_y$]

     \addplot [line width=0.5mm,smooth,black]  table [x expr=\thisrowno{0},y expr=\thisrowno{1}/600]{alphae_kin.txt};
     \addplot [line width=0.5mm,smooth,black,dashed]  table [x expr=\thisrowno{0},y expr=\thisrowno{2}/600]{alphae_kin.txt};
     \addplot [line width=0.5mm,smooth,black,dash pattern={on 7pt off 2pt on 1pt off 3pt}]  table [x expr=\thisrowno{0},y expr=\thisrowno{3}/600]{alphae_kin.txt};
     \addplot [line width=0.5mm,smooth,black,dotted]  table [x expr=\thisrowno{0},y expr=\thisrowno{4}/600]{alphae_kin.txt};

     \addlegendentry{Path 1}
     \addlegendentry{Path 2}
     \addlegendentry{Path 3}
     \addlegendentry{Path 4}

    \end{axis}
    \end{tikzpicture}
\caption{Effective back stress in the absence of the cohesive zone for kinematic hardening with $E_t=E/20$.}
\label{fig:alpha_e}    
\end{figure}

In contrast to the mode I loaded crack, the difference in the dissipation density is negligible when comparing an isotropic and a kinematic hardening material for a mode II loaded crack. Figure~\ref{fig:dissi_all_II} shows results along the four paths shown in Fig. \ref{fig:p_zone}b. For path 1, the dissipation density is slightly higher for isotropic hardening, whereas for path 2, 3, and 4 the energy dissipation density is slightly higher for the kinematic model. Thus, for the mode II crack, differences are marginal and no distinct region is observed where one model clearly dominates. 
\begin{figure}[!tb]
        \centering        
        \begin{subfigure}{0.45\textwidth}
        \centering
\resizebox{\linewidth}{!}{        
        \begin{tikzpicture}
    \begin{axis}[
        legend pos=north east,
        grid,
        xmin= -2,   xmax=9,
	    ymin= -0.2, ymax=2,
	    legend cell align={left},
        xlabel=$x_1/R_p$,
        ylabel=$w/(\Gamma/R_p)$]
        
     \addlegendimage{empty legend}
     \addlegendentry{\hspace{-.6cm}\textbf{\underline{Path 1}}}
   
     \addplot [line width=0.5mm,smooth,black,dashed]  table [x expr=\thisrowno{0},y expr=\thisrowno{1}/15.0796]{dissipation_iso_II.txt};
     \addlegendentry{Iso}
     \addplot [line width=0.5mm,stack plots=y,smooth,black]  table [x expr=\thisrowno{0},y expr=\thisrowno{1}/15.0796]{dissipation_kin_II.txt};
     \addlegendentry{Kin}
    
     \addplot [line width=0.5mm,stack plots=y, stack dir=minus,smooth,black,dotted]  table [x expr=\thisrowno{0},y expr=\thisrowno{1}/15.0796]{dissipation_iso_II.txt};
     \addlegendentry{Diff}
    
    \end{axis}
    \end{tikzpicture}
}
        \end{subfigure}
        \begin{subfigure}{0.45\textwidth}
        \centering
        \resizebox{\linewidth}{!}{
        \begin{tikzpicture}
    \begin{axis}[
        legend pos=north east,
        grid,
        xmin= -2,   xmax=9,
	    ymin= -0.2, ymax=2,
	    legend cell align={left},
        xlabel=$x_1/R_p$,
        ylabel=$w/(\Gamma/R_p)$]
        
     \addlegendimage{empty legend}
     \addlegendentry{\hspace{-.6cm}\textbf{\underline{Path 2}}}
     
     \addplot [line width=0.5mm,smooth,black,dashed]  table [x expr=\thisrowno{0},y expr=\thisrowno{2}/15.0796]{dissipation_iso_II.txt};
     \addlegendentry{Iso}
     \addplot [line width=0.5mm,stack plots=y,smooth,black]  table [x expr=\thisrowno{0},y expr=\thisrowno{2}/15.0796]{dissipation_kin_II.txt};
     \addlegendentry{Kin}
    
     \addplot [line width=0.5mm,stack plots=y, stack dir=minus,smooth,black,dotted]  table [x expr=\thisrowno{0},y expr=\thisrowno{2}/15.0796]{dissipation_iso_II.txt};
     \addlegendentry{Diff}
    
    \end{axis}
    \end{tikzpicture}
    }
        \end{subfigure}
        \begin{subfigure}{0.45\textwidth}
        \centering
        \resizebox{\linewidth}{!}{
        \begin{tikzpicture}
    \begin{axis}[
        legend pos=north east,
        grid,
        xmin= -2,   xmax=9,
	    ymin= -0.2, ymax=2,
	    legend cell align={left},
        xlabel=$x_1/R_p$,
        ylabel=$w/(\Gamma/R_p)$]
     
     \addlegendimage{empty legend}
     \addlegendentry{\hspace{-.6cm}\textbf{\underline{Path 3}}}
     
     \addplot [line width=0.5mm,smooth,black,dashed]  table [x expr=\thisrowno{0},y expr=\thisrowno{3}/15.0796]{dissipation_iso_II.txt};
     \addlegendentry{Iso}
     \addplot [line width=0.5mm,stack plots=y,smooth,black]  table [x expr=\thisrowno{0},y expr=\thisrowno{3}/15.0796]{dissipation_kin_II.txt};
     \addlegendentry{Kin}
    
     \addplot [line width=0.5mm,stack plots=y, stack dir=minus,smooth,black,dotted]  table [x expr=\thisrowno{0},y expr=\thisrowno{3}/15.0796]{dissipation_iso_II.txt};
     \addlegendentry{Diff}
   
  \end{axis}
    \end{tikzpicture}
    }
        \end{subfigure}
        \begin{subfigure}{0.45\textwidth}
        \centering
        \resizebox{\linewidth}{!}{
        \begin{tikzpicture}
    \begin{axis}[
        legend pos=north east,
        grid,
        xmin= -2,   xmax=9,
	    ymin= -0.2, ymax=2,
	    legend cell align={left},
        xlabel=$x_1/R_p$,
        ylabel=$w/(\Gamma/R_p)$]
     
     \addlegendimage{empty legend}
     \addlegendentry{\hspace{-.6cm}\textbf{\underline{Path 4}}}
     
     \addplot [line width=0.5mm,smooth,black,dashed]  table [x expr=\thisrowno{0},y expr=\thisrowno{4}/15.0796]{dissipation_iso_II.txt};
     \addlegendentry{Iso}
     \addplot [line width=0.5mm,stack plots=y,smooth,black]  table [x expr=\thisrowno{0},y expr=\thisrowno{4}/15.0796]{dissipation_kin_II.txt};
     \addlegendentry{Kin}
    
     \addplot [line width=0.5mm,stack plots=y, stack dir=minus,smooth,black,dotted]  table [x expr=\thisrowno{0},y expr=\thisrowno{4}/15.0796]{dissipation_iso_II.txt};
     \addlegendentry{Diff}
    
    \end{axis}
    \end{tikzpicture}
    }
        \end{subfigure}
        \caption{Mode II energy dissipation density in the absence of the cohesive zone for isotropic and kinematic hardening, and the difference between the models, with $E_t=E/20$.} 
        \label{fig:dissi_all_II}
\end{figure}

The effective back stress along the four paths in Fig. \ref{fig:p_zone}b is shown in Fig.~\ref{fig:alpha_II}. Here it is seen that the yield surface translates in the stress space until the crack tip is reached and remains stationary (constant $\alpha_e$) behind the crack. That is, no reversed plasticity is predicted for the mode II loaded crack, effectively eliminating one source of additional energy dissipation in the kinematic hardening model relative to the mode I analysis. In addition, the non-proportional loading is expected to be much less severe for the mode II crack, rationalizing the similar steady-state fracture toughness predictions obtained with isotropic and kinematic hardening models, respectively, (recall Fig. \ref{fig:shield_II}).
\begin{figure}
\centering
\begin{tikzpicture}
    \begin{axis}[
        legend style={font=\tiny},
        legend pos=north east,
        grid,
        xmin= -2,   xmax=9,
	    legend cell align={left},
        xlabel=$x_1/R_p$,
        ylabel=$\alpha_e/\sigma_y$]

     \addplot [line width=0.5mm,smooth,black]  table [x expr=\thisrowno{0},y expr=\thisrowno{1}/600]{alphae_kin_II.txt};
     \addplot [line width=0.5mm,smooth,black,dashed]  table [x expr=\thisrowno{0},y expr=\thisrowno{2}/600]{alphae_kin_II.txt};
     \addplot [line width=0.5mm,smooth,black,dash pattern={on 7pt off 2pt on 1pt off 3pt}]  table [x expr=\thisrowno{0},y expr=\thisrowno{3}/600]{alphae_kin_II.txt};
     \addplot [line width=0.5mm,smooth,black,dotted]  table [x expr=\thisrowno{0},y expr=\thisrowno{4}/600]{alphae_kin_II.txt};

     \addlegendentry{Path 1}
     \addlegendentry{Path 2}
     \addlegendentry{Path 3}
     \addlegendentry{Path 4}

    \end{axis}
    \end{tikzpicture}
\caption{Effective back stress in the absence of the cohesive zone for kinematic hardening with $E_t=E/20$.} 
\label{fig:alpha_II}    
\end{figure}

\section{Concluding remarks}
\label{S:conclusion}
A steady-state framework combined with a cohesive zone model has been developed for the purpose of studying the difference in cracks propagating in either isotropic or kinematic hardening materials. The study focuses on the plastic zone size, the evolution of the strain energy dissipation density, and the shielding ratio for mode I, mode II and mode I/II cracks. The main findings are:

\begin{itemize}
\item The shielding ratio is generally largest for the kinematic hardening material compared to an isotropic hardening material. The effect is most significant for a mode I crack whereas the effect is very limited for a mode II crack. Under mixed mode loading conditions, the shielding ratio falls between that of the two pure modes.

\item The active plastic zone for a mode I crack is slightly larger for a kinematic hardening material in the central half of the active plastic zone, whereas the isotropic hardening material has a larger plastic zone in the exterior half of the active plastic zone, for identical far-field loading. 
For the mode II crack, there is no significant difference in the plastic zone shapes.

\item The largest difference in energy dissipation density between the isotropic and kinematic hardening material for a mode I crack is observed close to the crack face, behind the crack tip, in the region of reverse loading. 
For the mode II crack, a very small difference in the dissipation density is observed. The dissipation density is slightly larger close to the crack face for isotropic hardening whereas further away from the crack face the kinematic model becomes dominant. The difference is, however, very small.\\

\end{itemize}

The main sources of the larger energy dissipation for kinematic hardening is attributed to the stronger path dependence associated with a larger curvature of the yield surface (sensitivity to non-proportional loading) and the reverse loading prone to initiate plasticity sooner than for the isotropic hardening material (smaller yield surface).
The large differences observed in mode I conditions between isotropic and kinematic predictions imply that fracture toughness estimations from R-curve modelling are very conservative. This could have important implications in damage tolerant design in the aerospace or energy sectors, among others.

\section{Acknowledgement}
\label{S:ack}
The work is financially supported by The Danish Council for Independent Research in the project ``New Advances in Steady-State Engineering Techniques", grant DFF-4184-00122. E. Mart\'{i}nez-Pa\~{n}eda additionally acknowledges financial support from the Ministry of Economy and Competitiveness of Spain through
grant MAT2014-58738-C3 and the People Programme (Marie Curie Actions) of the European Union’s Seventh Framework Programme (FP7/2007-2013) under REA grant agreement n◦ 609405 (COFUNDPostdocDTU).





\bibliographystyle{elsarticle-harv}
\bibliography{mybibfile}

\begin{thebibliography}{34}
\expandafter\ifx\csname natexlab\endcsname\relax\def\natexlab#1{#1}\fi
\expandafter\ifx\csname url\endcsname\relax
  \def\url#1{\texttt{#1}}\fi
\expandafter\ifx\csname urlprefix\endcsname\relax\def\urlprefix{URL }\fi

\bibitem[{Andersen et~al.(2019)Andersen, Felter, and
  Nielsen}]{Andersen_et_al_2018}
Andersen, R.~G., Felter, C.~L., Nielsen, K.~L., 2019. Micro-mechanics based
  cohesive zone modeling of full scale ductile plate tearing: From initiation
  to steady-state. (submitted for publication).

\bibitem[{Ashby(1970)}]{ashby1970}
Ashby, M.~F., 1970. {The deformation of plastically non-homogeneous materials}.
  Philosophical Magazine 21~(170), 399--424.

\bibitem[{Cao and Evans(1989)}]{cao1989a}
Cao, H.~C., Evans, A.~G., 1989. An experimental-study of the
  fracture-resistance of bimaterial interfaces. Mechanics of Materials 7~(4),
  295--304.

\bibitem[{Cleveringa et~al.(2000)Cleveringa, {Van der Giessen}, and
  Needleman}]{Cleveringa2000}
Cleveringa, H., {Van der Giessen}, E., Needleman, A., 2000. {A discrete
  dislocation analysis of mode I crack growth}. Journal of the Mechanics and
  Physics of Solids 48~(6–7), 1133--1157.

\bibitem[{de~Formanoir et~al.(2017)de~Formanoir, Brulard, Vives, Martin, Prima,
  Michotte, Riviere, Dolimont, and Godet}]{formanoir2017a}
de~Formanoir, C., Brulard, A., Vives, S., Martin, G., Prima, F., Michotte, S.,
  Riviere, E., Dolimont, A., Godet, S., 2017. A strategy to improve the
  work-hardening behavior of ti-6al-4v parts produced by additive
  manufacturing. Materials Research Letters 5~(3), 201--208.

\bibitem[{Dean and Hutchinson(1980)}]{key01}
Dean, R.~H., Hutchinson, J.~W., 1980. Quasi-static steady crack growth in
  small-scale yielding. Fracture Mechanics: Twelfth Conference, ASTM STP700,
  American Society for Testing and Materials, 383--405.

\bibitem[{del Busto et~al.(2017)del Busto, Betegon, and
  Mart\'{i}nez-Pa\~{n}eda}]{busto2017a}
del Busto, S., Betegon, C., Mart\'{i}nez-Pa\~{n}eda, E., 2017. A cohesive zone
  framework for environmentally assisted fatigue. Engineering Fracture
  Mechanics 185, 210--226.

\bibitem[{Freund and Hutchinson(1985)}]{Freund1985}
Freund, L.~B., Hutchinson, J.~W., 1985. {High strain-rate crack growth in
  rate-dependent plastic solids}. Journal of the Mechanics and Physics of
  Solids 33~(2), 169--191.

\bibitem[{Jiang et~al.(2010)Jiang, Wei, Smith, Hutchinson, and
  Evans}]{Jiang2010}
Jiang, Y., Wei, Y., Smith, J.~R., Hutchinson, J.~W., Evans, A.~G., 2010. {First
  principles based predictions of the toughness of a metal/oxide interface}.
  International Journal of Materials Research 101, 1--8.

\bibitem[{Juul et~al.(2017{\natexlab{a}})Juul, Nielsen, and
  Niordson}]{Juul2016b}
Juul, K.~J., Nielsen, K.~L., Niordson, C.~F., 2017{\natexlab{a}}. Steady-state
  crack growth in single crystals under mode i loading. J. Mech. Phys. Solids
  101, 209--222.

\bibitem[{Juul et~al.(2017{\natexlab{b}})Juul, Nielsen, and
  Niordson}]{Juul2016a}
Juul, K.~J., Nielsen, K.~L., Niordson, C.~F., 2017{\natexlab{b}}. Steady-state
  numerical modeling of size effects in micron scale wire drawing. J. Manuf.
  Process. 25, 163--171.

\bibitem[{Kumar and Curtin(2007)}]{Kumar2007}
Kumar, S., Curtin, W.~A., 2007. {Crack interaction with microstructure}.
  Materials Today 10~(9), 34--44.

\bibitem[{Landis et~al.(2000)Landis, Pardoen, and Hutchinson}]{Landis2000}
Landis, C.~M., Pardoen, T., Hutchinson, J.~W., 2000. {Crack velocity dependent
  toughness in rate dependent materials}. Mechanics of Materials 32~(11),
  663--678.

\bibitem[{Lemaitre and Chaboche(1990)}]{Lemaitre1990}
Lemaitre, J., Chaboche, J.-L., 1990. {Mechanics of solid materials}. Cambridge
  University Press, New York.

\bibitem[{Mart\'{i}nez-Pa\~{n}eda et~al.(2017)Mart\'{i}nez-Pa\~{n}eda, del
  Busto, and Betegon}]{Paneda2017a}
Mart\'{i}nez-Pa\~{n}eda, E., del Busto, S., Betegon, C., 2017. Non-local
  plasticity effects on notch fracture mechanics. Theor. Appl. Fract. Mech. 92,
  276--287.

\bibitem[{Mart{\'{i}}nez-Pa{\~{n}}eda and Fleck(2018)}]{JAM2018}
Mart{\'{i}}nez-Pa{\~{n}}eda, E., Fleck, N.~A., 2018. {Crack growth resistance
  in metallic alloys: the role of isotropic versus kinematic hardening}.
  Journal of Applied Mechanics 85, 11002 (6 pages).

\bibitem[{Mart{\'{i}}nez-Pa{\~{n}}eda and Niordson(2016)}]{IJP2016}
Mart{\'{i}}nez-Pa{\~{n}}eda, E., Niordson, C.~F., 2016. {On fracture in finite
  strain gradient plasticity}. International Journal of Plasticity 80,
  154--167.

\bibitem[{Mear and Hutchinson(1985)}]{Mear1985}
Mear, M.~E., Hutchinson, J.~W., 1985. {Influence of yield surface curvature on
  flow localization in dilatant plasticity}. Mechanics of Materials 4,
  395--407.

\bibitem[{Nielsen and Niordson(2012b)}]{key05}
Nielsen, K.~L., Niordson, C.~F., 2012b. Rate sensitivity of mixed mode
  interface toughness of dissimilar metallic materials: Studied at steady
  state. Int. J. Solids Struct. 49, 576--583.

\bibitem[{Nielsen et~al.(2012a)Nielsen, Niordson, and Hutchinson}]{Kim2012}
Nielsen, K.~L., Niordson, C.~F., Hutchinson, J.~W., 2012a. Strain gradient
  effects on steady-state crack growth in rate-dependent materials. Eng. Fract.
  Mech. 96, 61--71.

\bibitem[{Niordson(2001)}]{Niordson2001}
Niordson, C.~F., 2001. Analysis of steady-state ductile crack growth along a
  laser weld. Int. J. Frac. 111, 53--69.

\bibitem[{Segurado and LLorca(2004)}]{Segurado}
Segurado, J., LLorca, J., 2004. A new three-dimensional interface finite
  element to simulate fracture in composites. Int. J. Solids and Struct. 41,
  2977--2993.

\bibitem[{Suo et~al.(1993)Suo, Shih, and Varias}]{Suo1993}
Suo, Z., Shih, C.~F., Varias, A.~G., 1993. A theory for cleavage cracking in
  the presence of plastic flow. Acta Metall Mater 41, 1551--7.

\bibitem[{Tvergaard(1978)}]{tvergaard1978a}
Tvergaard, V., 1978. Effect of kinematic hardening on localized necking in
  biaxially stretched sheets. Int J Mech Sci 20~(9), 651--658.

\bibitem[{Tvergaard(2001)}]{tvergaardbook}
Tvergaard, V., 2001. Plasticity and creep in structural materials. Polyteknisk
  Kompendie.

\bibitem[{Tvergaard(2002)}]{tvergaard2002a}
Tvergaard, V., 2002. Theoretical investigation of the effect of plasticity on
  crack growth along a functionally graded region between dissimilar
  elastic-plastic solids. Engineering Fracture Mechanics 69~(14-16),
  1635--1645.

\bibitem[{Tvergaard(2010)}]{tvergaard2010a}
Tvergaard, V., 2010. Effect of pure mode i, ii or iii loading or mode mixity on
  crack growth in a homogeneous solid. International Journal of Solids and
  Structures 47~(11-12), 1611--1617.

\bibitem[{Tvergaard and Hutchinson(1992)}]{tvergaard1992a}
Tvergaard, V., Hutchinson, J.~W., 1992. The relation between crack-growth
  resistance and fracture process parameters in elastic plastic solids. Journal
  of the Mechanics and Physics of Solids 40~(6), 1377--1397.

\bibitem[{Tvergaard and Hutchinson(1993)}]{tvergaard1993a}
Tvergaard, V., Hutchinson, J.~W., 1993. The influence of plasticity on
  mixed-mode interface toughness. Journal of the Mechanics and Physics of
  Solids 41~(6), 1119--1135.

\bibitem[{Varias and Shih(1993)}]{varias1993a}
Varias, A.~G., Shih, C.~F., 1993. Quasi-static crack advance under a range of
  constraints - steady-state fields based on a characteristic length. Journal
  of the Mechanics and Physics of Solids 41~(5), 835--861.

\bibitem[{Wei and Hutchinson(1997)}]{Wei1997}
Wei, Y., Hutchinson, J.~W., 1997. Steady-state crack growth and work of
  fracture for solids characterized by strain gradient plasticity. J. Mech.
  Phys. Solids 45, 1253--73.

\bibitem[{Williams(1957)}]{Williams1957}
Williams, M.~L., 1957. On the stress distribution at the base of a stationary
  crack. J. App. Mech. 24, 109--114.

\bibitem[{Woelke et~al.(2015)Woelke, Shields, and Hutchinson}]{woelke2015a}
Woelke, P.~B., Shields, M.~D., Hutchinson, J.~W., 2015. Cohesive zone modeling
  and calibration for mode i tearing of large ductile plates. Engineering
  Fracture Mechanics 147, 293--305.

\bibitem[{Ziegler(1959)}]{ziegler1959a}
Ziegler, H., 1959. A modification of prager’s hardening rule. Quarterly of
  Applied Mathematics 17~(1), 55--65.

\end{thebibliography}





\end{document}